\documentclass[journal,10pt]{IEEEtran}
\usepackage{amssymb}
\usepackage{amsmath}
\usepackage{cite}
\usepackage{url}
\usepackage{xcolor}
\usepackage{cite,graphicx,amsmath,amssymb}
\usepackage{subfigure}
\usepackage{citesort}
\usepackage{fancyhdr}
\usepackage{mdwmath}
\usepackage{mdwtab}
\usepackage{caption}
\usepackage{amsthm}
\usepackage{setspace}
\usepackage{algorithm}
\usepackage{algorithmic}
\usepackage{makecell}
\usepackage{diagbox}
\usepackage{stfloats}
\usepackage{balance} 
\usepackage{graphicx}
\usepackage{epstopdf}
\usepackage{epsfig}

\newtheorem{theorem}{Theorem}

\newtheorem{lemma}{Lemma}

\newtheorem{corollary}{Corollary}

\allowdisplaybreaks
\makeatletter
\def\ScaleIfNeeded{
	\ifdim\Gin@nat@width>\linewidth \linewidth \else \Gin@nat@width
	\fi } \makeatother
 
\begin{document}
	\title{SE-EE Tradeoff in Pinching-Antenna Systems: Waveguide Multiplexing or Waveguide Switching?}
	\author{
		Guangyu~Zhu,
		Xidong~Mu,
		Li~Guo,
		Shibiao~Xu,
		Yuanwei Liu, \emph{Fellow, IEEE},
		Naofal Al-Dhahir, \emph{Fellow, IEEE}
		\thanks{Guangyu Zhu, Li Guo, and Shibiao Xu are with the School of Artificial Intelligence, Beijing University of Posts and Telecommunications, Beijing 100876, China (email:\{Zhugy, guoli, shibiaoxu\}@bupt.edu.cn).}
		\thanks{Xidong Mu is with the Centre for Wireless Innovation (CWI), Queen's University Belfast, Belfast, BT3 9DT, U.K. (e-mail:
			x.mu@qub.ac.uk).}
		\thanks{Yuanwei Liu is with the Department of Electrical and Electronic Engineering, The University of Hong Kong, Hong Kong. (e-mail:
			yuanwei@hku.hk).}
		\thanks{Naofal Al-Dhahir is with the Department of Electrical and Computer
			Engineering, The University of Texas at Dallas, Richardson, TX 75080 USA.
			(e-mail: aldhahir@utdallas.edu).}
	}
	\maketitle
	\begin{abstract}
		The spectral and energy efficiency (SE-EE) tradeoff in pinching-antenna systems (PASS) is investigated in this paper. In particular, two practical operating protocols, namely waveguide multiplexing (WM) and waveguide switching (WS), are considered. A multi-objective optimization problem (MOOP) is formulated to jointly optimize the baseband and pinching beamforming for maximizing the achievable SE and EE, which is then converted into a single-objective problem via the $\epsilon$-constraint method. For WM, the problem is decomposed within the alternating-optimization framework, where the baseband beamforming is optimized using the successive convex approximation, and the pinching beamforming is updated through the particle swarm optimization. For WS, due to the time-division transmission and interference-free nature, the pinching beamforming in each time slot is first adjusted to maximize the served user channel gain, followed by the baseband power allocation. Simulation results demonstrate that 1) PASS outperforms conventional antennas by mitigating large-scale path losses; 2) WS leads to a higher maximum achievable EE by activating a single RF chain, whereas WM yields a higher SE upper bound by serving all users concurrently; and 3) increasing the number of users substantially enhances SE under WM, whereas WS shows more pronounced benefits in low-signal-to-noise ratio regimes.
	\end{abstract}
	\begin{IEEEkeywords}
		Pinching-antenna systems, spectral efficiency, energy efficiency, waveguide multiplexing, waveguide switching.
	\end{IEEEkeywords}
\section{Introduction}
Spectral efficiency (SE) has long been a key performance indicator in wireless communication systems. In addition, with the increasing demand for sustainable network operation, energy efficiency (EE) has become equally critical. However, simultaneously improving SE and EE remains challenging because higher SE often requires increased transmit power or additional active hardware, which directly degrades EE. To alleviate this conflict, several flexible antenna technologies, such as reconfigurable intelligent surfaces (RISs) \cite{RIS_survey}, movable antennas \cite{Zhu_MA}, and fluid antennas \cite{Wong_Fluid}, have been proposed. These designs can enhance SE by reconfiguring the propagation environment or adjusting antenna positions, while their passive or low-power nature helps reduce energy consumption, thereby contributing to improved EE. Nevertheless, their flexibility and scalability remain limited: RISs lack baseband processing capability, whereas movable and fluid antennas are constrained by restricted motion ranges. As a result, their ability to jointly enhance SE and EE, especially in large-scale fading-dominated scenarios, is fundamentally restricted.
 
Recently, pinching‐antenna systems (PASS) have emerged as a promising architecture for future wireless networks \cite{Ding_PASS}.
By placing reconfigurable dielectric pinching antennas (PAs) along a dielectric waveguide, PASS enables flexible aperture shaping and dynamically adjustable line-of-sight transmission with extremely low propagation loss. PAs are typically composed of passive dielectric materials and therefore incur almost no additional power consumption, while their mobility allows PASS to effectively mitigate the large-scale fading. These characteristics make PASS inherently suitable for achieving both high SE and EE. Inspired by these advantages, the authors of \cite{Zhou_PASS} investigated the fundmental SE–EE tradeoff in PASS. However, it neglected the circuit power of activated RF chains and focused exclusively on the waveguide-multiplexing (WM) operating protocol, leaving the fundamental SE–EE behavior of PASS largely unexplored.

For multiple-waveguide PASS, WM is attractive because it serves multiple users simultaneously and fully exploits baseband beamforming to enhance the SE. However, WM requires all RF chains to be activated, which results in significant power consumption. To overcome this limitation, we introduce a waveguide-switching (WS) operating protocol in this paper, where only one waveguide is activated at the same time to serve a single user with all available PAs. This strategy reduces energy consumption and inter-user interference, while PAs from other inactive waveguides can be repositioned to the active waveguide to boost channel gain. Nevertheless, WS reduces each user’s available transmission time, especially when the number of users grows, which may limit its achievable SE. These contrasting characteristics naturally motivate a comprehensive comparative study of WM and WS to determine which protocol yields a more favorable SE-EE tradeoff, which forms the main focus of this work. 

\section{System Model and Problem Formulation}
\begin{figure}
	\setlength{\abovecaptionskip}{0cm}   
	\setlength{\belowcaptionskip}{0cm}   
	\setlength{\textfloatsep}{7pt}
	\centering
	\includegraphics[width=2.85in]{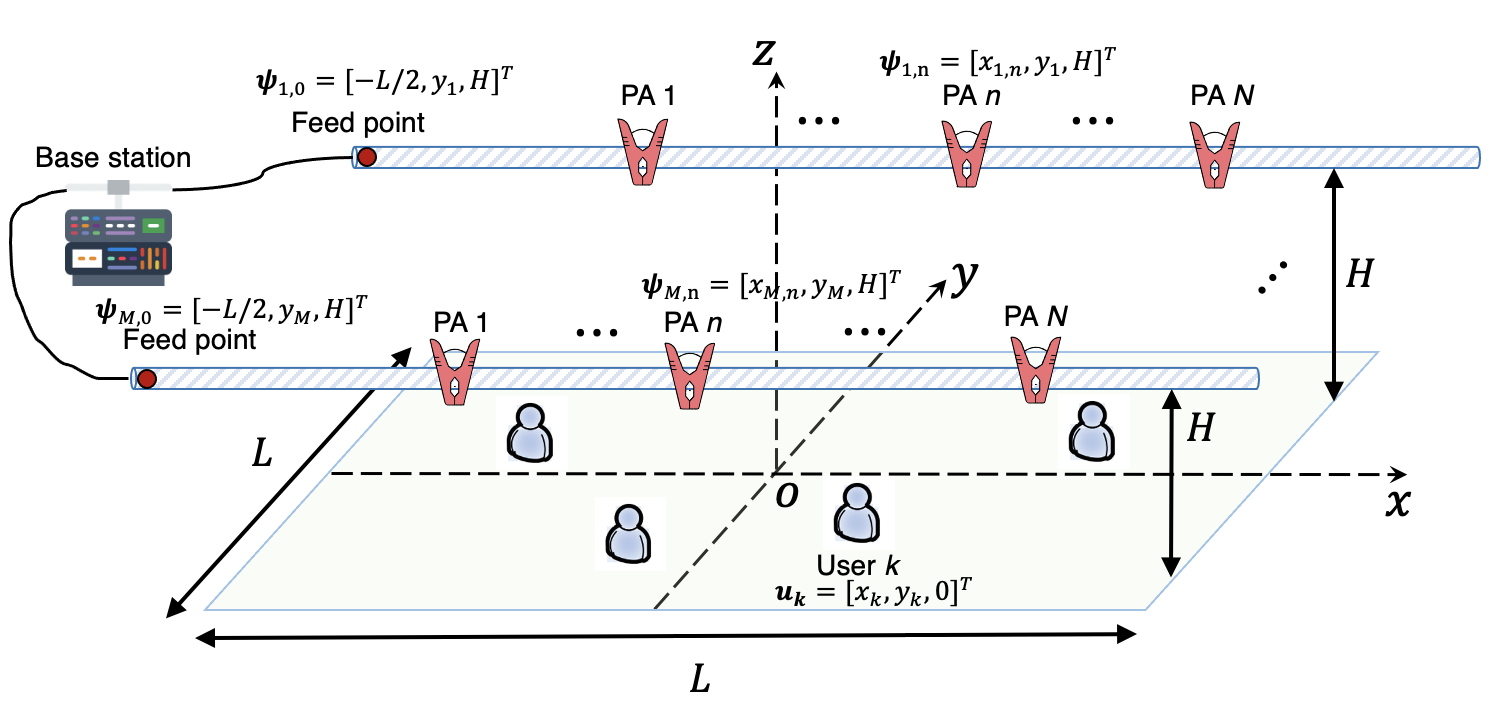}
	\caption{Illustration of a PASS-based wireless communication system.}
	\label{SW_system}
\end{figure}
As illustrated in Fig. \ref{SW_system}, we consider a PASS-based downlink wireless communication system, where a base station (BS) serves a set of $K$ single-antenna users, indexed by $\mathcal{K}=\{1,2,\cdots,K\}$, randomly located in a square area of size $L\times L$ centered at the origin. To enhance the transmission performance, the BS is equipped with $M$ parallel dielectric waveguides, each uniformly spaced by a fixed horizontal interval $d$, and all aligned along the $x$-axis at a uniform height of $H$ above the ground. For analytical clarity, we assume $M=K$, ensuring sufficient multiplexing gain to support $K$ users concurrently, which is commonly adopted in multiuser system analysis.
On each waveguide, $N$ PAs are deployed for flexible directional transmission. Let $\mathcal{M}=\{1,2,\cdots,M\}$ denote the set of waveguide indices and $\mathcal{N}=\{1,2,\cdots,N\}$ denote the set of PA indices per waveguide. The $n$-th PA on the $m$-th waveguide is located at the position $\boldsymbol{\psi}_{m,n}=[x_{m,n},y_m,H]^T$, where $-L/2 \leq x_{m,n} \leq L/2$, and $(.)^{T}$ denotes the transpose operation. Specifically, we assume $\boldsymbol{\psi}_{m,0}=[-L/2,y_m,H]^T$ as the reference feed point for the $m$-th waveguide, while the PAs are deployed in a successive order, meaning $x_{m,n} > x_{m,n-1}, \forall n \in \mathcal{N}$. In this case, the in-waveguide channel vector from the feed point to different PAs can be expressed as
\begin{align}
	\mathbf{e}(\mathbf{x}_m)\!\!=\!\!\bigg[\!e^{-j\frac{2\pi}{\lambda_g}\left\|\boldsymbol{\psi}_{m,0}\!-\boldsymbol{\psi}_{m,1}\right\|},\cdots\!,e^{-j\frac{2\pi}{\lambda_g}\left\|\boldsymbol{\psi}_{m,0}\!-\boldsymbol{\psi}_{m,N}\right\|}\bigg]^T\!,
\end{align}
where $\left\|\boldsymbol{\psi}_{m,0}-\boldsymbol{\psi}_{m,n}\right\|=(x_{m,n}+L/2)$ denotes the distance from the feed point to the $n$-th PA along the waveguide. The vector 
$\mathbf{x}_m=[x_{m,1},\cdots,x_{m,N}]^T$ specifies the $x$-coordinates of the $N$ PAs placed along the $m$-th waveguide. Besides, we note that $\lambda_g=\frac{\lambda}{n_{eff}}$, where $\lambda$ represents the wavelength and $n_{eff}$ denotes the effective refractive index of a dielectric waveguide.
	
Assuming that the $k$-th user is located at $\mathbf{u}_k=[x_k,y_k,0]^T$, the free-space channel vector between the $m$-th waveguide and the $k$-th user is given by
\begin{align}
	\widetilde{\mathbf{h}}_{k}(\mathbf{x}_m)\!=\!\bigg[\frac{\sqrt{\eta}e^{-j\frac{2\pi}{\lambda}\|\mathbf{u}_k-\boldsymbol{\psi}_{m,1}\|}}{\|\mathbf{u}_k-\boldsymbol{\psi}_{m,1}\|}\!,\cdots\!,\!\frac{\sqrt{\eta}e^{-j\frac{2\pi}{\lambda}\|\mathbf{u}_k-\boldsymbol{\psi}_{m,N}\|}}{\|\mathbf{u}_k-\boldsymbol{\psi}_{m,N}\|}\bigg]^T.
\end{align}
Here, $\sqrt{\eta}=\frac{c}{4\pi f_c}$ denotes the path loss coefficient where $c$ is the speed of light and $f_c$ is the carrier frequency. The corresponding wavelength is $\lambda=\frac{c}{f_c}$. Furthermore, $\|\mathbf{u}_k-\boldsymbol{\psi}_{m,n}\|=\sqrt{(x_k-x_{m,n})^2+(y_k-y_m)^2+H^2}$ denotes the distance from the $n$-th PA of the $m$-th waveguide to user $k$. As a consequence, the complete channel coefficient from the $m$-th waveguide to user $k$ can be expressed as follows
	\begin{align}
		h_{k}(\mathbf{x}_m)&\!=\!\widetilde{\mathbf{h}}^H_k(\mathbf{x}_m)\mathbf{e}(\mathbf{x}_m) \nonumber \\
		&\!=\!\sum^{N}_{n=1}\!\frac{\sqrt{\eta}e^{-2\pi j \left(\frac{1}{\lambda}\!\left\|\mathbf{u}_k-\boldsymbol{\psi}_{m,n}\right\|+\frac{1}{\lambda_g}\!\left\|\boldsymbol{\psi}_{m,0}\!-\boldsymbol{\psi}_{m,n}\right\|\right)}}{\|\mathbf{u}_k-\boldsymbol{\psi}_{m,n}\|}.
	\end{align}

\begin{figure}[t]
	\subfigure[Waveguide multiplexing (WM).]{\label{WM}
		\includegraphics[width= 3.0in]{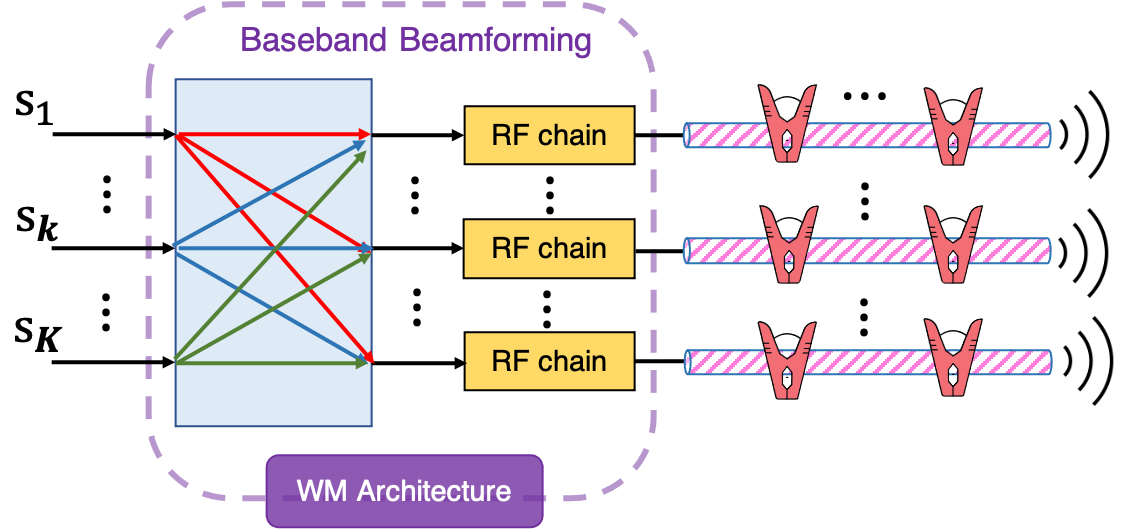}}
	\subfigure[Waveguide switching (WS).]{\label{WS}
		\includegraphics[width= 3.0in]{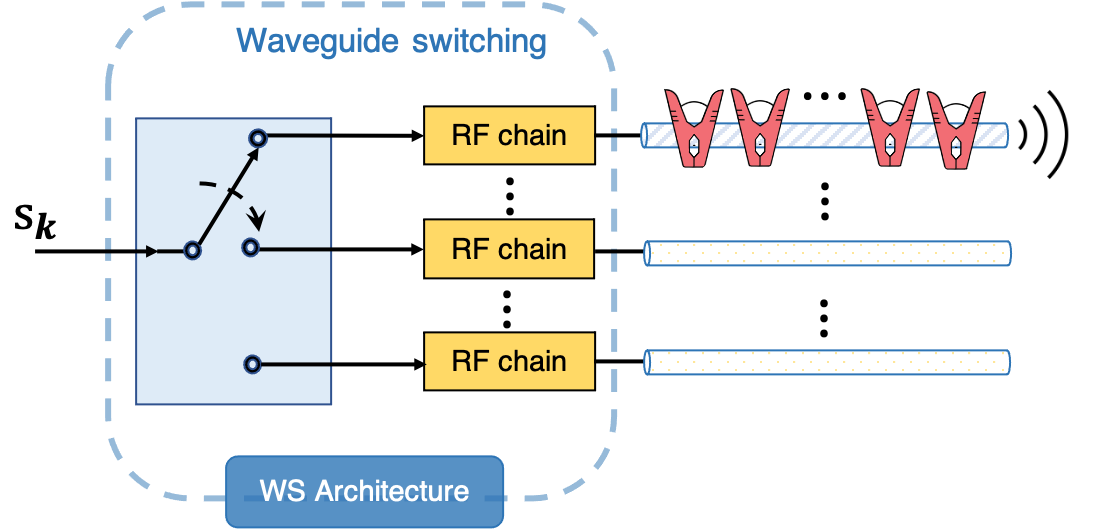}}
	\caption{\textcolor{black}{Illustration of two practical operating protocols.}}
	\label{Architecture}
\end{figure}
\subsection{Operating Protocols and Signal Transmission Models}
In this paper, we consider two practical operating protocols, namely WM and WS, as shown in Fig. \ref{Architecture}. 

\emph{1) Waveguide Multiplexing:} For WM, all waveguides are used concurrently for information delivery, as illustrated in Fig. \ref{WM}, and baseband signal processing is employed for multiplexing, which significantly enhances transmission efficiency. Consequently, the received signal at user $k$ is given by
\begin{align}
	y_k=\mathbf{h}_k\sum_{k=1}^K\mathbf{w}_ks_k+n_k,
\end{align}
where $\mathbf{h}_k=[h_k(\mathbf{x}_1),\cdots,h_k(\mathbf{x}_M)]^T$, and $\mathbf{w}_k$ denotes the information-bearing vector for user $k$. Besides, $s_k$, satisfying $\mathbb{E}\left[|s_k|^2\right]=1$, and $n_k \sim \mathcal{CN}(0, \sigma_k^2)$ denote the signal emitted from the baseband and additive white Gaussian noise at user $k$, respectively. 

Since the BS transmits information to all users simultaneously, each user is able to utilize the entire communication period. However, this inevitably leads to inter-user interference. Therefore, the achievable rate of user $k$ can be expressed as
\begin{align}
	R^{\textup{WM}}_k=\log_2\left(1+\frac{\big|\mathbf{h}_k\mathbf{w}_k\big|^2}{\sum_{i=1,i\neq k}^K\big|\mathbf{h}_k\mathbf{w}_i\big|^2+\sigma_k^2}\right).
\end{align}
The SE is calculated as the achievable rate of all users
\begin{align}
	f_{SE}(\mathbf{w}_k,\mathbf{x}_m)\!=\!\sum_{k=1}^K\!\log_2\!\left(\!1\!+\!\frac{\big|\mathbf{h}_k\mathbf{w}_k\big|^2}{\sum_{i=1,i\neq k}^K\!\big|\mathbf{h}_k\mathbf{w}_i\big|^2\!\!+\!\sigma_k^2}\right).
\end{align}
Furthermore, the EE is defined as the ratio of SE to the total power consumption and is given by
\begin{align}
	f_{EE}(\mathbf{w}_k,\mathbf{x}_m)=\frac{f_{SE}(\mathbf{w}_k,\mathbf{x}_m)}{\sum_{k=1}^K\|\mathbf{w}_k\|^2+KP_{RF}},
\end{align}
where $P_{RF}$ denotes the power consumption for each RF chain.

\emph{2) Waveguide Switching:}
As depicted in Fig. \ref{WS}, under the WS protocol, only one waveguide is activated for data transmission during each cycle, which allows activation of a single RF chain and thus reduces power consumption. Meanwhile, PAs associated with the idle waveguides can be repositioned to the active one, offering enhanced design flexibility. According to this transmission characteristic, the received signal at user $k$ is given by
\begin{align}
	y_k=\sqrt{\frac{P_k}{N}}h_k(\mathbf{x}_m)s_k+n_k.
\end{align}
Here, the equal power model is adopted and $P_k$ denotes the transmit power of the BS to user $k$. In particular, the selection of $m$ is inherently determined by the position of user $k$. 

Under the time-division communication feature for WS, each user can transmit without interference only during its assigned time slot. Assuming that all users are allocated equal time resources\cite{Qian_PASS}, the achievable rate of user $k$ can be expressed as
\begin{align}
	R^{\textup{WS}}_k=\frac{1}{K}\log_2\left(1+\frac{P_k\big|h_k(\mathbf{x}_m)\big|^2}{N\sigma_k^2}\right).
\end{align}
Accordingly, the achievable SE for WS is given by
\begin{align}
	f_{SE}(P_k,\mathbf{x}_m)=\sum_{k=1}^{K}\frac{1}{K}\log_2\left(1+\frac{P_k\big|h_k(\mathbf{x}_m)\big|^2}{N\sigma_k^2}\right).
\end{align}
Furthermore, the EE for WS is given by
\begin{align}
	f_{EE}(P_k,\mathbf{x}_m)=\frac{Kf_{SE}(P_k,\mathbf{x}_m)}{\sum_{k=1}^KP_k+KP_{RF}}.
\end{align}

\subsection{Problem Formulation}
To explore the achievable tradeoff between SE and EE in the system, we consider a joint optimization of the baseband beamforming $\{\mathbf{w}_k\}$ or transmit power $P_k$ together with the pinching beamforming $\{\mathbf{x}_m\}$, subject to the system constraints, aiming to simultaneously maximize both SE and EE. Noting that EE can be expressed as the ratio of SE to total power consumption, the problem can equivalently be reformulated as a multi-objective optimization problem (MOOP) that seeks to maximize SE while minimizing the overall power consumption. Furthermore, since the RF chain power $P_{RF}$ is constant, it is ignored in the optimization. As a result, the MOOP can be formally expressed as follows
\begin{subequations}\label{P_MOOP}
	\begin{align}
		&\label{P_MO_C0}\ \max_{\{\mathbf{w}_k\},\{\mathbf{x}_m\}} f_{SE}\left(\mathbf{w}_k,\mathbf{x}_m\right) \bigg/ \max_{\{P_k\},\{\mathbf{x}_m\} }f_{SE}\left(P_k,\mathbf{x}_m\right) \\
		&\label{P_MO_C1}\ \min_{\{\mathbf{w}_k\},\{\mathbf{x}_m\}} \sum_{k=1}^K\|\mathbf{w}_k\|^2\bigg/ \min_{\{P_k\},\{\mathbf{x}_m\} } \sum_{k=1}^K P_k\\
		&\label{P_MO_WM_C2}\ \quad {\rm s.t.} \ \ \sum_{k=1}^K \|\mathbf{w}_k\|^2 \leq P_{\max}, \\
		&\label{P_MO_WS_C3}\  \quad \quad \quad P_k \leq P_{\max}, \forall k \in\mathcal{K}, \\
		&\label{P_MO_C4}\  \quad \quad \quad R^{\textup{X}}_k \geq \gamma_k,\forall k \in\mathcal{K}, \textup{X} \in \{\textup{WM},\textup{WS}\}, \\
		&\label{P_MO_C5} \quad \quad \quad -L/2<x_{m,n}<L/2, \forall n \in \mathcal{N}, \forall m \in \mathcal{M},\\
		&\label{P_MO_C6}\ \quad \quad \quad  x_{m,n}-x_{m,n-1}\geq \Delta, \forall n \in \mathcal{N}, \forall m \in \mathcal{M}.
	\end{align}
\end{subequations}
Here, constraints \eqref{P_MO_WM_C2} and \eqref{P_MO_WS_C3} impose the transmit power budgets for the WM and WS protocols, respectively, where $P_{\max}$ denotes the maximum allowable transmit power. In addition, $\gamma_k$ in \eqref{P_MO_C4}  denotes the minimum quality-of-service (QoS) requirement for each user. Moreover, constraints \eqref{P_MO_C5} and \eqref{P_MO_C6} regulate the PA positions, with $\Delta$ being the minimum spacing to prevent coupling effect. Note that $\mathcal{N}$ differs across protocols. In WM, each waveguide contains $N$ PAs, so $\mathcal{N}$ runs from $1$ to $N$. By contrast, WS effectively aggregates all PAs onto one active waveguide per cycle making $\mathcal{N}$ runs from $1$ to $MN$.

\section{Proposed Solutions}
To address the MOOP, we adopt the $\epsilon$-constraint method, which converts the MOOP into a single-objective power minimization problem subject to a minimum SE requirement. Specifically, for each target SE threshold $\epsilon_{SE}$, we solve a corresponding single-objective optimization problem (SOOP). The detailed formulations of the SOOP for the WM and WS protocols are presented separately in the following
\begin{subequations}\label{P_WM_SOOP}
	\begin{align}
		&\label{P_WM_SO_C0} \min_{\{\mathbf{w}_k\},\{\mathbf{x}_m\}} \sum_{k=1}^K\|\mathbf{w}_k\|^2\\
		&\label{P_WM_SO_C1}  \quad \quad {\rm s.t.}\ f_{SE}(\mathbf{w}_k,\mathbf{x}_m) \geq \epsilon_{SE},\\
		&\label{P_WM_SO_C2} \quad \quad \quad \ \ \eqref{P_MO_WM_C2}, \eqref{P_MO_C4},\eqref{P_MO_C5}, \eqref{P_MO_C6}.
	\end{align}
\end{subequations}
\vspace{-0.5cm}
\begin{subequations}\label{P_WS_SOOP}
	\begin{align}
		&\label{P_WS_SO_C0} \min_{\{P_k\},\{\mathbf{x}_m\}} \sum_{k=1}^K P_k \\
		&\label{P_WS_SO_C1}  \quad \quad {\rm s.t.}\ f_{SE}(P_k,\mathbf{x}_m) \geq \epsilon_{SE},\\
		&\label{P_WS_SO_C2} \quad \quad \quad \ \ \eqref{P_MO_WS_C3}, \eqref{P_MO_C4},\eqref{P_MO_C5}, \eqref{P_MO_C6}.
	\end{align}
\end{subequations}
\subsection{Proposed Solution for WM}
Due to the strong coupling between the pinching beamforming variables $\{\mathbf{x}_m\}$ and the baseband beamforming vectors $\{\mathbf{w}_k\}$, the original problem \eqref{P_WM_SOOP} is difficult to solve directly. To address this issue, we adopt an alternating optimization (AO) framework that decomposes the original problem into two subproblems, pinching beamforming design and baseband beamforming design, which are solved in an iterative manner.

\emph{1) Baseband Beamforming Optimization:} Given fixed PA positions $\{\mathbf{x}_m\}$, we define $\mathbf{W}_k\triangleq\mathbf{w}_k\mathbf{w}_K^H \in \mathbb{C}^{M\times M}$, under which the baseband beamforming subproblem becomes a standard semidefinite programming problem as follows
\begin{subequations}\label{P_WM_BB}
	\begin{align}
		&\label{P_WM_BB_C0} \min_{\{\mathbf{W}_k\}}\ \ \sum_{k=1}^K\mathrm{Tr}(\mathbf{W}_k)\\
		&\label{P_WM_BB_C1}  {\rm s.t.}\ \sum_{k=1}^K\mathrm{Tr}(\mathbf{W}_k) \leq P_{\max},\\
		&\label{P_WM_BB_C2}  \ \ \quad \mathrm{Rank}(\mathbf{W}_k)=1, \forall k \in \mathcal{K},\\
		&\label{P_WM_BB_C3}  \ \ \quad \mathbf{W}_k \succeq 0, \forall k \in \mathcal{K},\\
		&\label{P_WM_BB_C4}  \quad \ \  \log_2\left(1\!+\!\frac{\mathrm{Tr}\left(\mathbf{h}_k\mathbf{W}_k\mathbf{h}^H_k\right)}{\sum^K_{i=1,i\neq k}\mathrm{Tr}\left(\mathbf{h}_k\mathbf{W}_i\mathbf{h}^H_k\right)\!+\!\sigma^2_k}\right) \geq \gamma_k, \\
		&\label{P_WM_BB_C5}  \ \sum^K_{k=1}\!\log_2\left(\!1\!+\!\frac{\mathrm{Tr}\left(\mathbf{h}_k\mathbf{W}_k\mathbf{h}^H_k\right)}{\sum^K_{i=1,i\neq k}\mathrm{Tr}\left(\mathbf{h}_k\mathbf{W}_i\mathbf{h}^H_k\right)\!+\!\sigma^2_k}\right) \!\geq\! \epsilon_{SE}.		
	\end{align}
\end{subequations}
However, the achievable rate is non-convex in $\{\mathbf{W}_i\}^K_{i=1,i\neq k}$ and the rank-one constraints on 
$\{\mathbf{W}_k\}$ introduce additional non-convexity, making the problem still difficult to solve in its current form. To handle this issue, we adopt the successive convex approximation (SCA) method, where the achievable rate is linearized at a given point $\{\mathbf{W}^{(l)}_i\}^K_{i=1,i\neq k}$ via a first-order Taylor expansion to obtain a convex lower bound as
\begin{align}
	R^{\textup{WM}}_k\!\!\!& \geq\! \log_2\left(\sum_{k=1}^K \mathrm{Tr}\left(\mathbf{h}_k\mathbf{W}_k\mathbf{h}^H_k\right)+\sigma^2_k\right) \nonumber \\
	&-\log_2\left(\sum^K_{i=1,i\neq k}\mathrm{Tr}\left(\mathbf{h}_k\mathbf{W}^{(l)}_i\mathbf{h}^H_k\right)+\sigma^2_k\right)\nonumber \\
	&-\frac{\sum^K_{i=1,i\neq k}\!\mathrm{Tr}\left(\mathbf{h}_k\left(\mathbf{W}_i\!-\!\mathbf{W}^{(l)}_i\right)\mathbf{h}^H_k\right)}{\left(\sum^K_{i=1,i\neq k}\mathrm{Tr}\left(\mathbf{h}_k\mathbf{W}^{(l)}_i\mathbf{h}^H_k\right)\!+\!\sigma^2_k\right)\ln2} \!\triangleq \!\widetilde{R}^{\textup{WM}}_k.
\end{align}

Furthermore, by invoking Lemma 2 in \cite{Zhu_PASS}, the non-convex rank-one constraint \eqref{P_WM_BB_C2} can be safely dropped without altering the optimal solution. We therefore solve its semidefinite relaxation, which is equivalent in optimal value. After these steps, the problem \eqref{P_WM_BB} is reformulated as follows
\begin{subequations}\label{P_WM_BB_Relax}
	\begin{align}
		&\label{P_WM_BB_R_C0} \min_{\{\mathbf{W}_k\}}\ \  \sum_{k=1}^K\mathrm{Tr}(\mathbf{W}_k)\\
		&\label{P_WM_BB_R_C1}  \ {\rm s.t.}\ \widetilde{R}^{\textup{WM}}_k \geq \gamma_k, \forall k \in\mathcal{K},\\
		&\label{P_WM_BB_R_C2}  \ \quad \ \  \sum^K_{k=1} \widetilde{R}^{\textup{WM}}_k \geq \epsilon_{SE}, \\
		&\label{P_WM_BB_R_C3} \ \quad \ \  \eqref{P_WM_BB_C1}, \eqref{P_WM_BB_C3}.
	\end{align}
\end{subequations}
This is a standard convex problem, which can be solved by existing solvers like CVX \cite{cvx}.

\emph{2) Pinching Beamforming Design:} With $\{\mathbf{w}_k\}$ fixed, this subproblem has no independent objective and reduces to a feasibility-oriented PA position search as follows
\begin{subequations}\label{P_WM_PB}
	\begin{align}
		&\label{P_WM_PB_C0} \textup{Find} \quad \ \ \mathbf{x}_m\\
		&\label{P_WM_PB_C1}  \ {\rm s.t.}\  \eqref{P_MO_C4},\eqref{P_MO_C5}, \eqref{P_MO_C6}, \eqref{P_WM_SO_C1}.
	\end{align}
\end{subequations}

To efficiently address this subproblem, we extend the particle swarm optimization (PSO) algorithm in \cite{Zhu_PASS} from the two-waveguide case to the general $M$-waveguide scenario. We then initialize a swarm of particles of size $I$ as 
\begin{align}
	\mathbf{X}_i^{(0)}=[x^{(0)}_{i,1},\!\cdots\!,x^{(0)}_{i,N},\!\cdots\!,x^{(0)}_{i,MN-N+1},\!\cdots\!,x^{(0)}_{i,MN}]^T,
\end{align}
where $x_{i,(m-1)N+n}$ represents the $x$-coordinate of the $n$-th PA on the $m$-th waveguide in the $i$-th particle. Correspondingly, we define the initial velocity of these particle swarms as 
\begin{align}
	\mathbf{V}_i^{(0)}=[v^{(0)}_{i,1},\!\cdots\!,v^{(0)}_{i,N},\!\cdots\!,v^{(0)}_{i,MN-N+1},\!\cdots\!,v^{(0)}_{i,MN}]^T.
\end{align}
Following the PSO principles, each particle updates its position iteratively based on its personal best $\mathbf{X}_{i,p}$ and the global best $\mathbf{X}_{g}$ of the swarm, as given by the update equation below
\begin{gather}
	\label{velocity}\mathbf{V}^{(t+1)}_{i}\!\!=\alpha\mathbf{V}^{(t)}_i\!+\!c_1\beta_1(\mathbf{X}_{i,p}\!-\!\mathbf{X}^{(t)}_i)\!+\!c_2\beta_2(\mathbf{X}_{g}\!-\!\mathbf{X}^{(t)}_i),\\
	\label{position}\mathbf{X}^{(t+1)}_i=\mathbf{X}^{(t)}_i+\mathbf{V}^{(t+1)}_i,
\end{gather}
where $t$ is the iteration index, $\alpha$ is the inertia weight, $c_1$ and $c_2$ denote the learning factors, and $\beta_1$ and $\beta_2$ denote random numbers in [0,1] adding stochasticity. To accelerate convergence and guide the search toward feasible and high-quality solutions, we adopt the following fitness function
\begin{align} \label{fitness}
	\mathcal{F}(\mathbf{X}_i)=\mathcal{R}(\mathbf{X}_i)-\xi\mathcal{P}_1(\mathbf{X}_i)-\eta\mathcal{P}_2(\mathbf{X}_i).
\end{align}
Here, $\mathcal{R}$ denotes the system SE to be maximized, while $\mathcal{P}_1$ and $\mathcal{P}_2$  correspond to the achievable performance and PA position constraints in the original problem, incorporated as penalty terms with coefficients $\xi$ and $\eta$, respevtively.

\subsection{Proposed Solution for WS}
For WS, user transmissions are orthogonal in time, and hence interference among users is completely avoided. Under fixed QoS requirements, minimizing the BS transmit power becomes equivalent to maximizing the effective channel gain for each user. As a result, problem \eqref{P_WS_SOOP} can be equivalently decomposed into $K$ independent channel-gain maximization subproblems as follows
\begin{subequations}\label{Problem_WS_channel_gain}
	\begin{align}
		&\label{P_WS_CG_C0}\  \max_{\{\mathbf{x}_m\}}\quad \bigg| \sum^N_{n=1}\frac{e^{-j\theta_{m,n}}}{\sqrt{(x_{m,n}-x_k)^2+D^2_k}}\bigg|\\ 
		&\label{P_SW_CG_C1}\quad {\rm s.t.} \quad \eqref{P_MO_C5},\eqref{P_MO_C6}.
	\end{align}
\end{subequations}
where $\theta_{m,n}={2\pi \left(\frac{1}{\lambda}\left\|\mathbf{u}_k-\boldsymbol{\psi}_{m,n}\right\|+\frac{1}{\lambda_g}\left\|\boldsymbol{\psi}_{m,0}-\boldsymbol{\psi}_{m,n}\right\|\right)}$, $D_k=\sqrt{(y_k-y_{m})^2+H^2}$. 

To solve this problem, the optimal waveguide for user $k$ can be selected as the one whose position $y_m$
is closest to the user’s $y$-coordinate, i.e., minimizing $|y_k-y_m|$. Once the working waveguide is determined, the optimal positions of the PAs on that waveguide can be obtained according to \cite[\textbf{Lemma 2}]{Xu_PASS}. 

With the optimal PA positions $\textbf{x}^*_m$ determined for each user $k$, the original problem can be reduced to a power allocation problem as
 \begin{subequations}\label{P_WS_power_allocation}
 	\begin{align}
 		&\label{P_WS_PA_C0} \min_{P_k,\{\mathbf{x}_m\}} \sum_{k=1}^KP_k\\
 		&\label{P_WS_PA_C1}  \quad \quad {\rm s.t.}\ f_{SE}(P_k,\mathbf{x}^*_m) \geq \epsilon_{SE},\\
 		&\label{P_WS_PA_C2} \quad \quad \quad \ \ \eqref{P_MO_WS_C3}, \eqref{P_MO_C4}.
 	\end{align}
 \end{subequations}
The resulting problem is a linear programming problem and can be directly solved via CVX \cite{cvx}.

\section{Numerical Results}
In this section, numerical results are presented to evaluate the SE-EE tradeoff in PASS under both the WM and WS protocols. The simulation parameters are set as follows. All waveguides have a height of $H=3$ m with an inter-waveguide spacing of $d=1$ m, and the minimum PA spacing to avoid coupling is $\Delta=\frac{\lambda}{2}$. The communication region length is $L=10$ m, the communication carrier frequency is $f_c=28$ GHz, the maximum transmit power at the BS is $P_{\max}=100$ mW, the power consumption per RF chain is $P_{RF} = 31.6$ mW \cite{RF_power}, the noise power is $\sigma^2_k=-90$ dBm, and each user’s QoS requirement is $\gamma_k= 1$ bit/s/Hz.

For comparison, we consider a baseline scheme, termed $\textbf{conventional antenna}$, where the BS deployed $M$ RF chains at positions $\left[-L/2, y_m, H\right]^T$, each connected to an $N$-antenna array and operating with hybrid beamforming. The SE-EE tradeoff curves are obtained using the $\epsilon$-constraint method. By sweeping the achievable SE with appropriate step sizes, the minimum transmit power for each SE is computed, and the corresponding EE is calculated as SE divided by this power, resulting in the tradeoff curves shown in the following figures.

\begin{figure}
	\setlength{\abovecaptionskip}{0cm}   
	\setlength{\belowcaptionskip}{0cm}   
	\setlength{\textfloatsep}{7pt}
	\centering
	\includegraphics[width=3.0in]{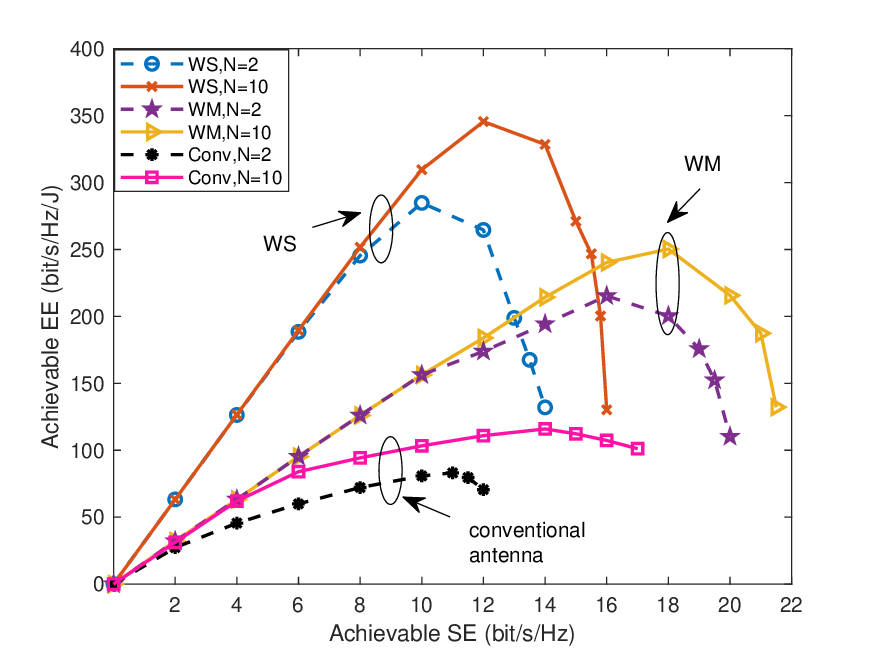}
	\caption{Achievable SE-EE tradeoff versus different $N$.}
	\label{res_DW_PAs}
\end{figure}
Fig. \ref{res_DW_PAs} illustrates the SE-EE tradeoff for different antenna numbers $N$ under WS, WM, and conventional antenna architectures. In all cases, EE increases with SE in the low-SE regime but decreases when SE exceeds a certain threshold. This is because at low SE, the required transmit power is relatively low compared with the circuit power of RF chains, so improving SE leads to a significant EE gain. At higher SE, however, the BS transmit power becomes dominant, and the additional power required to further increase SE grows more significant than the SE improvement itself, resulting in an EE decline. Besides, both the WM and WS protocols under PASS achieve higher performance than the conventional antenna. This is due to the flexible PA positioning, which mitigates large-scale fading and enhances transmission efficiency. Regarding the comparison between WM and WS, WS attains higher maximum EE, as only a single RF chain is activated at any time, thereby reducing the circuit power consumption. In contrast, WM achieves a higher maximum SE, since all users can occupy the full communication period, whereas in WS each user is allocated only a fraction $1/K$ of the period. Moreover, increasing $N$ expands the achievable SE-EE tradeoff region for all schemes. This is because a larger antenna array provides more design degrees of freedom (DoFs). The performance improvement is particularly notable for WS, as the additional DoFs introduced by each waveguide accumulate within a single active waveguide, yielding extra performance gains.

\begin{figure}
	\setlength{\abovecaptionskip}{0cm}   
	\setlength{\belowcaptionskip}{0cm}   
	\setlength{\textfloatsep}{7pt}
	\centering
	\includegraphics[width=3.0in]{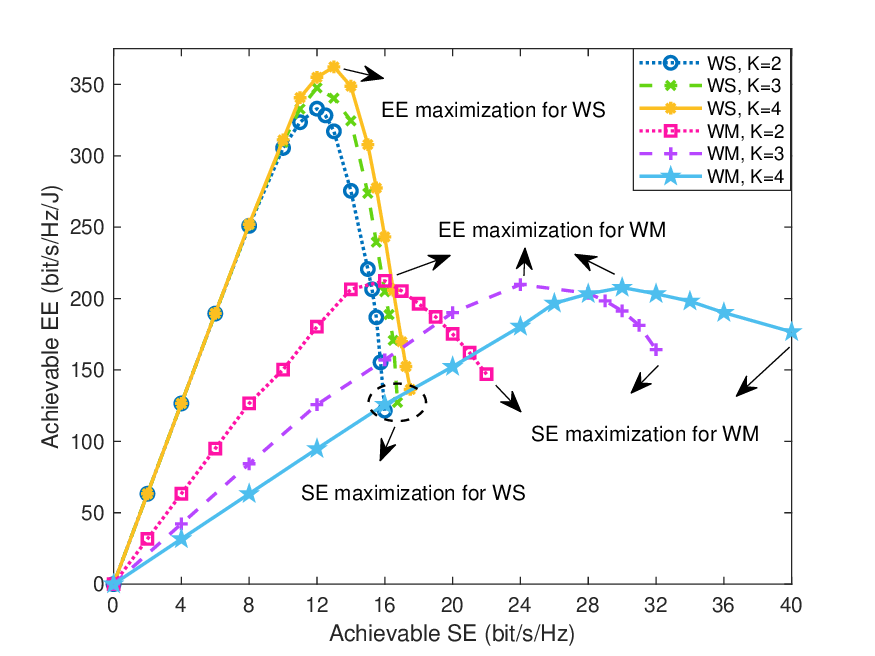}
	\caption{Achievable SE-EE tradeoff versus different users.}
	\label{res_user}
\end{figure}
Fig. \ref{res_user} illustrates the SE-EE tradeoff versus the number of users $K$. As $K$ increases, all schemes exhibit an expanded tradeoff region, while WM expands more prominently. This is because more activated waveguides provide WM with higher baseband beamforming DoF, enabling more effective signal enhancement and interference suppression. In contrast, WS is limited by the $1/K$ time allocation per user. Nevertheless, in the low-SE regime, WS achieves higher EE than WM as $K$ increases. This is because WS activates only a single RF chain, whereas WM requires multiple RF chains and thus incurs higher circuit power consumption.

\section{Conclusions}
This paper investigated the SE-EE tradeoff in the PASS under both WM and WS protocols. A MOOP was established for joint beamforming design, and the problem was subsequently solved using the $\epsilon$-constraint method. For WM, alternating optimization with SCA and PSO was employed, while for WS, pinching beamforming in each time slot was optimized to maximize user channel gain, followed by baseband power allocation. Simulation results showed that WS can achieve higher EE, while WM can attain greater SE. Moreover, increasing the number of users further boosts SE in WM, whereas WS maintains advantages in low-SE regimes. These findings highlight the practical potential of PASS for sepctral- and energy-efficient wireless communications.
\bibliographystyle{IEEEtran} 
\bibliography{pass.bib}

\begin{thebibliography}{10}
\providecommand{\url}[1]{#1}
\csname url@samestyle\endcsname
\providecommand{\newblock}{\relax}
\providecommand{\bibinfo}[2]{#2}
\providecommand{\BIBentrySTDinterwordspacing}{\spaceskip=0pt\relax}
\providecommand{\BIBentryALTinterwordstretchfactor}{4}
\providecommand{\BIBentryALTinterwordspacing}{\spaceskip=\fontdimen2\font plus
\BIBentryALTinterwordstretchfactor\fontdimen3\font minus
  \fontdimen4\font\relax}
\providecommand{\BIBforeignlanguage}[2]{{%
\expandafter\ifx\csname l@#1\endcsname\relax
\typeout{** WARNING: IEEEtran.bst: No hyphenation pattern has been}%
\typeout{** loaded for the language `#1'. Using the pattern for}%
\typeout{** the default language instead.}%
\else
\language=\csname l@#1\endcsname
\fi
#2}}
\providecommand{\BIBdecl}{\relax}
\BIBdecl

\bibitem{RIS_survey}
Y.~Liu, X.~Liu, X.~Mu, T.~Hou, J.~Xu, M.~Di~Renzo, and N.~Al-Dhahir,
  ``Reconfigurable intelligent surfaces: Principles and opportunities,''
  \emph{{IEEE} Commun. Surv. Tut.}, vol.~23, no.~3, pp. 1546--1577, 3rd Quart.
  2021.

\bibitem{Zhu_MA}
L.~Zhu, W.~Ma, and R.~Zhang, ``Modeling and performance analysis for movable
  antenna enabled wireless communications,'' \emph{{IEEE} Trans. Wireless
  Commun.}, vol.~23, no.~6, pp. 6234--6250, Jun. 2024.

\bibitem{Wong_Fluid}
K.-K. Wong, A.~Shojaeifard, K.-F. Tong, and Y.~Zhang, ``Fluid antenna
  systems,'' \emph{{IEEE} Trans. Wireless Commun.}, vol.~20, no.~3, pp.
  1950--1962, Mar. 2021.

\bibitem{Ding_PASS}
Z.~Ding, R.~Schober, and H.~Vincent~Poor, ``Flexible-antenna systems: A
  pinching-antenna perspective,'' \emph{{IEEE} Trans. Commun.}, vol.~73,
  no.~10, pp. 9236--9253, Oct. 2025.

\bibitem{Zhou_PASS}
Z.~Zhou, Z.~Wang, and Y.~Liu, ``Spectral and energy efficiency tradeoff for
  pinching-antenna systems,'' [Online].
  Available:\url{https://arxiv.org/abs/2510.25192}.

\bibitem{Qian_PASS}
M.~Qian, X.~Mu, L.~You, and M.~Matthaiou, ``Pinching-antenna-based
  communications: Spectral efficiency analysis and deployment strategies,''
  [Online]. Available:\url{https://arxiv.org/abs/2507.14831}, 2025.

\bibitem{Zhu_PASS}
G.~Zhu, X.~Mu, L.~Guo, S.~Xu, Y.~Liu, and N.~Al-Dhahir, ``Pinching-antenna
  systems ({PASS})-enabled secure wireless communications,'' \emph{{IEEE}
  Trans. Commun.}, pp. 1--1, early access, Oct. 2025,
  doi:10.1109/TCOMM.2025.3621084.

\bibitem{cvx}
M.~Grant and S.~Boyd, ``{CVX}: {MATLAB} software for disciplined convex
  programming, version 2.1,'' [Online]. Available:\url{http://cvxr.com/cvx},
  2014.

\bibitem{Xu_PASS}
Y.~Xu, Z.~Ding, and G.~K. Karagiannidis, ``Rate maximization for downlink
  pinching-antenna systems,'' \emph{{IEEE} Wireless Commun. Lett.}, vol.~14,
  no.~5, pp. 1431--1435, May 2025.

\bibitem{RF_power}
M.~R. Castellanos, S.~Yang, C.-B. Chae, and R.~W. Heath, ``Embracing
  reconfigurable antennas in the tri-hybrid {MIMO} architecture for {6G} and
  beyond,'' \emph{{IEEE} Trans. Commun.}, pp. 1--1, 2025.

\end{thebibliography}
\end{document}